\documentclass[psf,proceedingpaper,accept,pdftex,moreauthors]{Definitions/mdpi} 

\firstpage{1} 
\pubvolume{12}
\issuenum{1}
\articlenumber{0}
\pubyear{2025}
\copyrightyear{2025}
\externaleditor{Geert Verdoolaege} 
\datereceived{ } 
\daterevised{ } 
\dateaccepted{ } 
\datepublished{ } 
\hreflink{https://doi.org/} 

\Title{Understanding Exoplanet Habitability: A Bayesian ML Framework for Predicting Atmospheric Absorption Spectra $^{\dagger}$}

\TitleCitation{Understanding Exoplanet Habitability: A Bayesian ML Framework for Predicting Atmospheric Absorption Spectra}

\Author{Vasuda Trehan $^{1,}$*$^{,\ddagger}$\orcidA{}, Kevin H. Knuth $^{2,}$*$^{,\ddagger}$\orcidB{} and M. J. Way $^{3,4,}$*$^{,\ddagger}$\orcidC{}}

\AuthorNames{Vasuda Trehan, Kevin H. Knuth and M. J. Way}

\AuthorCitation{Trehan, V.; Knuth, K.H.; Way, M.J.}

\address{%
$^{1}$ \quad Department of Information Science, University at Albany, Albany, NY 12222, USA\\
$^{2}$ \quad Department of Physics, University at Albany, Albany, NY 12222, USA\\
$^{3}$ \quad NASA Goddard Institute for Space Studies, New York, NY 10025, USA\\
$^{4}$ \quad Theoretical Astrophysics, Department of Physics \& Astronomy, Uppsala University, 75105 Uppsala, Sweden}

\corres{Correspondence: kknuth@albany.edu (K.H.K.)} 

\firstnote{Presented at the 43rd International Workshop on Bayesian Inference and Maximum Entropy Methods in Science and Engineering, Ghent, Belgium, 1–5 July 2024.}  
\secondnote{These authors contributed equally to this work.}

\abstract{The evolution of space technology in recent years, fueled by advancements in computing such as Artificial Intelligence (AI) and machine learning (ML), has profoundly transformed our capacity to explore the cosmos. Missions like the James Webb Space Telescope (JWST) have made information about distant objects more easily accessible, resulting in extensive amounts of valuable data.  As part of this work-in-progress study, we are working to create an atmospheric absorption spectrum prediction model for exoplanets. The eventual model will be based on both collected observational spectra and synthetic spectral data generated by the ROCKE-3D general circulation model (GCM) developed by the climate modeling program at NASA's Goddard Institute for Space Studies (GISS). In this initial study, spline curves are used to describe the bin heights of simulated atmospheric absorption spectra as a function of one of the values of the planetary parameters.  Bayesian Adaptive Exploration is then employed to identify areas of the planetary parameter space for which more data are needed to improve the model. The resulting system will be used as a forward model so that planetary parameters can be inferred given a planet's atmospheric absorption spectrum. This work is expected to contribute to a better understanding of exoplanetary properties and general exoplanet climates and habitability.
}


\begin{document}

\section{Introduction}

In the expanding field of exoplanet studies, we aim to develop a machine learning system that will predict exoplanetary atmospheric absorption spectra based on a set of approximately 30 planetary parameters (PPs). This system will be trained on both real and synthetic data.  This data includes spectra recorded from terrestrial planets within our solar system, along with synthetic spectra modeled using the ROCKE-3D~\cite{Way+etal:2017} general circulation model (GCM) at NASA's Goddard Institute for Space Studies (GISS). In this paper, as an early step towards building our final framework, we present a proof-of-concept implementation of a simplified version of this framework. This system promises to enhance our understanding of exoplanetary properties, as well as general planetary habitability.

By predicting the atmospheric absorption spectra based on given planetary parameters, this machine learning system will effectively act as a predictive forward model, which can then be used to evaluate likelihood functions for a Bayesian inference engine that will infer probable exoplanet parameters from observed atmospheric absorption spectra. Much of the initial training data will be sourced from recorded spectra from Earth, Venus, Mars, and Titan, as well as synthetic spectra generated by simulation studies conducted using 3D climate models of the Archean and Proterozoic Earth, as well as similar past and present climate models of Venus and Mars~\cite{LUVOIR:2019, Schmid+etal:2022, Way+etal:2020}.

\begin{figure}[H]
 \begin{center}
\includegraphics[width=1.0\columnwidth]{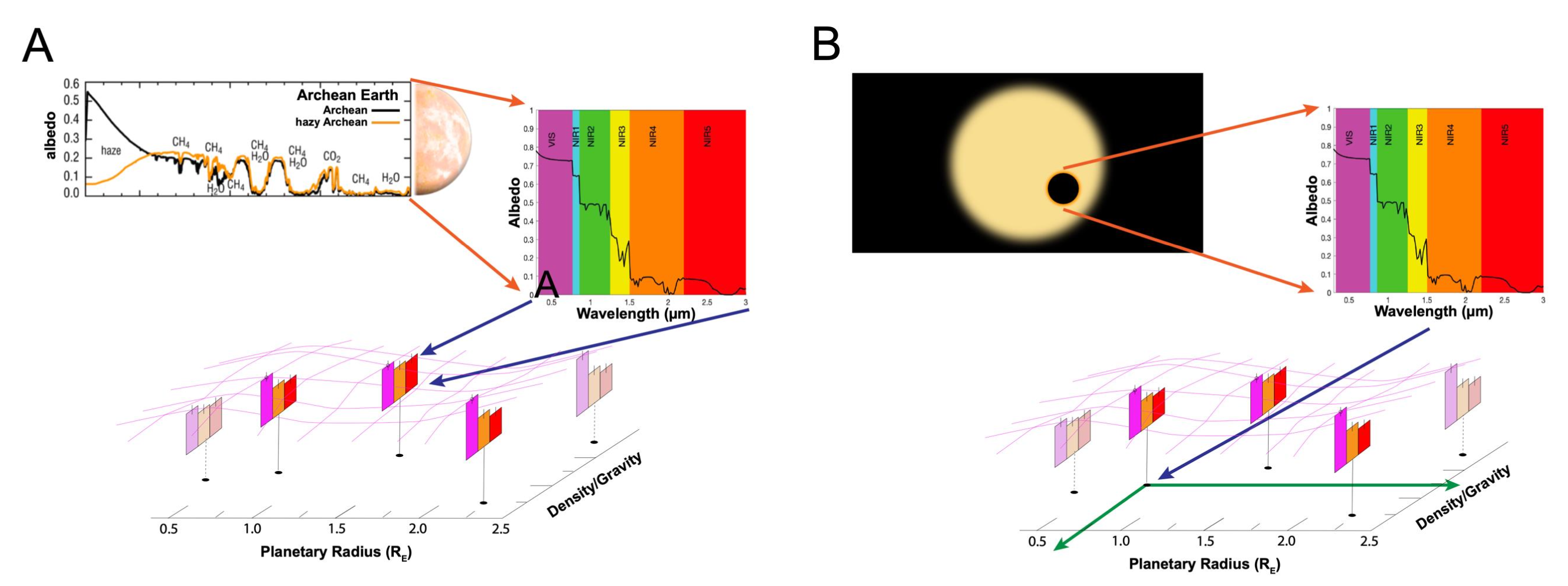}
\end{center}
\caption{(\textbf{A}). The machine learning system will be initially trained on recorded present spectra and historic synthetic spectra generated by ROCKE-3D GCM simulations at NASA's GISS. (\textbf{B}). Once trained, the system can serve as a predictive forward model that will enable a Bayesian inference engine to estimate planetary parameters from recorded exoplanetary spectra.\label{fig-1}}
\end{figure}

The machine learning system will be trained (Figure \ref{fig-1}A) so that it will be able to predict atmospheric absorption spectra from planetary parameters.  Once this has been accomplished, the system’s ability to predict spectra will make it useful for evaluating likelihood functions in a Bayesian inference engine that will then estimate planetary parameters based on atmospheric spectra, as illustrated in Figure \ref{fig-1}B.

\begin{figure}[H]

\includegraphics[width=0.8\columnwidth]{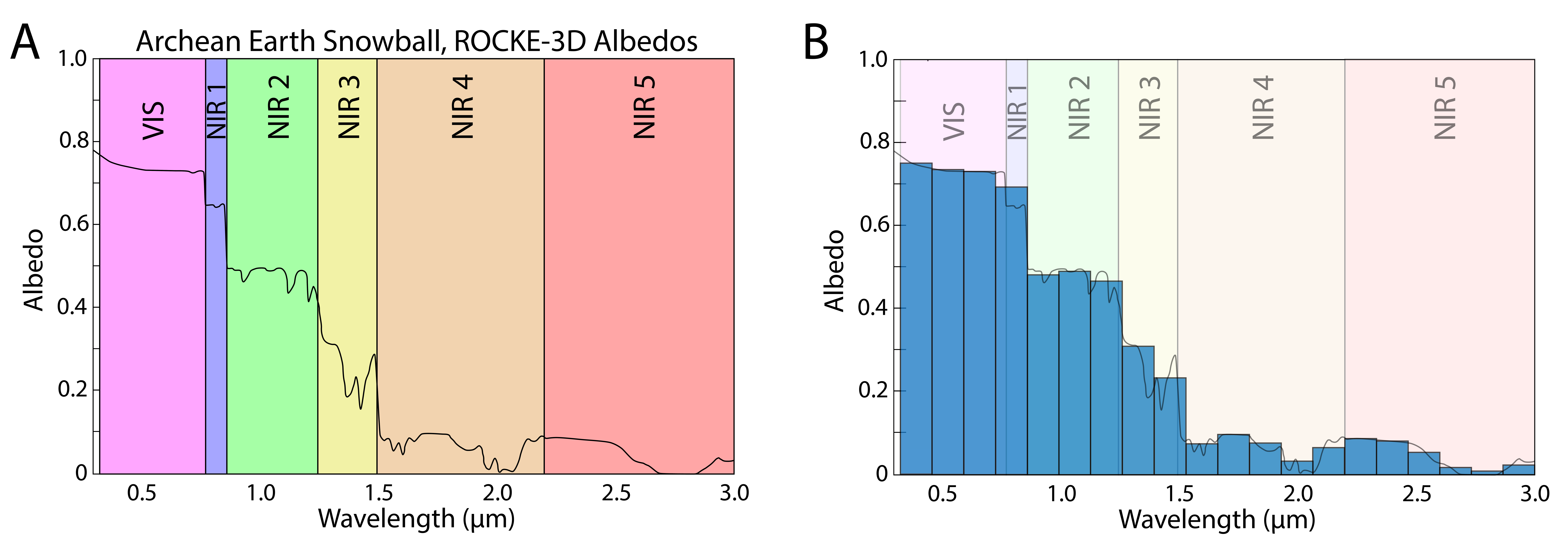}

\caption{(\textbf{A}). A synthetic atmospheric absorption spectrum, ranging from the visible (VIS) range through to near-infrared (NIR), of Archean snowball Earth generated using ROCKE-3D~\cite{Way+etal:2023}. (\textbf{B}). A summary of the spectrum as a set of 20 discrete bins. \label{fig-2}}
\end{figure}

The atmospheric spectra will be modeled as discrete bins consisting of approximately 20 bins spanning the visible (VIS) to near-infrared (NIR) (Figure \ref{fig-2}). Each planet is characterized by approximately 30 parameters, such as planetary radius, orbital radius, stellar classification, day-side temperature, and oxygen content, to name a few. This complex modeling involves predicting multiple (20) scalar values (bin values of spectral intensity) in a 30-dimensional planetary parameter space.

As a proof-of-concept, we are working with one planetary parameter, limiting the problem to one dimension and focusing on just six spectral bins. At this stage, we are employing spline models to address the design of a prototype model while being aware that we will need to adopt a model that will scale to 30 dimensions. Our current focus is to demonstrate the use of spline interpolation and Bayesian Adaptive Exploration in a low-dimensional setting before scaling it to a higher-dimensional model. As the dimensionality increases, so will the complexity; hence, designing an efficient model and establishing a viable methodology are of prime importance.

This research discusses the various methodologies that could be implemented to design the model, as well as their possible limitations and improvements for further research. This paper is divided into sections based on the following format: Section \ref{sec2} reviews related work and provides the contextual background.  Section \ref{sec3} delves into the problem of modeling spectra. Section \ref{sec4} discusses the other potential approaches, followed by Section \ref{sec5} that discusses the methodology, explaining the mathematical formulation and approach. Section \ref{sec6} presents the results and limitations of the current methodology, and finally, Section \ref{sec7} offers concluding statements.

\section{Related Work and Context}\label{sec2}
The launch of advanced space-based observatories, such as the James Webb Space Telescope (JWST), has enabled the acquisition of medium-resolution atmospheric spectral data for exoplanets. This influx of detailed observational data has accelerated progress in atmospheric characterization and highlighted the need for scalable and computationally efficient analysis frameworks. McDonald and Batalha~\cite{MacDonald+Batalha:2023} provide a comprehensive overview of the current atmospheric retrieval techniques, many of which use Bayesian inference incorporated using machine learning to enhance the parameter estimation and spectral fitting~\cite{MacDonald+Batalha:2023}. Alternatively, detailed examinations with forward modeling approaches, which help us understand how light interacts with planetary atmospheres to reveal their chemical composition, can be achieved using tools like Tau-REx~\cite{waldmann2015taurex}, NEMESIS~\cite{irwin2008nemesis}, and CHIMERA~\cite{line2013chimeara, 
 Rengel+Adamczewski:2023}.~
These frameworks treat atmospheric retrieval as an inverse problem, using computationally intensive, precomputed forward models to infer atmospheric properties.

Recent efforts have focused on the integration of general circulation models (GCMs)~\cite{GCM:web} to improve both scalability and realism. For example, Aura-3D~\cite{Nixon+Madhusudhan:2022} performs transmission spectrum retrieval using full 3D atmospheric simulations. Machine learning has been used to enhance the OASIS framework~\cite{Tahseen+etal:2024} by reducing the number of parameters. In another approach, where GCMs like MITgcm~\cite{MITgcm:latest} are being adapted for exoplanet studies~\cite{Ranftl+vonderLinden:2021}, their use in surrogate modeling remains limited due to the complexity and high computational cost of simulating exoplanet spectra. Most current approaches tend to focus on isolated parts of the retrieval pipeline, either forward modeling or parameter inference, rather than offering a fully integrated solution.

To address these limitations and gaps, our work aims to develop a scalable forward surrogate model capable of predicting atmospheric absorption spectra from approximately 30~planetary parameters. This work constitutes a generic proof-of-concept involving a single planetary parameter and spline interpolation. By adopting a spline model (which works well with a single parameter) with machine learning, we have developed a predictive system that fits seamlessly into a Bayesian inference framework. This enables parameter estimation and supports active data acquisition via Bayesian Adaptive Exploration\cite{Loredo:2004}. Unlike traditional inverse methods, our approach generates complete spectral outputs while also quantifying uncertainty. A key strength of our model lies in its use of ROCKE-3D, a terrestrial GCM originally designed for Earth climate studies and later adapted for exoplanet research~\cite{Way+etal:2017} to produce physically consistent training data. This hybrid strategy enables a more holistic and efficient exploration of exoplanetary atmospheres, moving beyond the limitations of conventional retrieval frameworks.

\section{Modeling Spectra}\label{sec3}

The long-term goal of this project revolves around characterizing a planet using approximately 30 parameters and relating these parameters to the planet’s atmospheric absorption spectra.  Each spectrum will be described using a set of approximately 20 discrete bins, like a histogram~\cite{Knuth:2019}. Figure \ref{fig-2}A provides an example of an expected spectrum. This spectrum is derived from a ROCKE-3D simulation of an Archean Snowball Earth~\cite{Way+etal:2023}, exhibiting six distinct albedo bands used in the GCM: the visible (VIS) range, near-infrared (NIR), and subsequent bands up to a wavelength of $3\,\mu\mathrm{m}$.  A detailed synthetic spectrum will be summarized as a set of 20 bins (Figure \ref{fig-2}B).

The objective of the model is to predict the heights of these 20 bins on a bar graph (Figure \ref{fig-2}B) for a given planet based on its planetary parameter values, where the height of each bin is expected to be a function of up to 30 planetary parameters. Considering each parameter on its individual dimension, a multidimensional model is created. The heights, $h$, of the 20 bins can be expressed as follows:
\begin{equation}
    h_1 = F_1(p_1, p_2, \ldots, p_{30}); \; \ldots; \; h_{b} = F_{b}(p_1, p_2, \ldots, p_{30})  \; \ldots; \; h_{20} = F_{20}(p_1, p_2, \ldots, p_{30})
\end{equation}
where $h_b$ signifies the height of the $b$-th bin, $p_{1}$ through $p_{30}$ denote the 30 planetary parameters, and $F_b$ represents the functional dependence of the $b$-th bin's height on the planetary parameters. 

However, this task presents challenges, and to address these challenges, we simplify the problem in this paper by avoiding the complexity of considering all 30 planetary parameters. We instead considered a simple model of 6 bins instead of 20 bins and focused on the problem in one dimension with one planetary parameter. We model each of the six functions, $F_1, \ldots, F_b, \ldots, F_6$, as a Piecewise Cubic Hermite Interpolating Polynomial (PCHIP)~\cite{Fritsch+Carlson:1980}, which is a specific kind of spline curve~\cite{Ahlberg:1967}.  The PCHIP is characterized by a set of $K$ spline knots, each defined by the planetary parameter value $x$ and the spectra bin height $f$ at that planetary parameter value, $\{(x_i, f_i), (x_{i+1}, f_{i+1}), \ldots, (x_{x+K-1}, f_{x+K-1})\}$, such that~\cite{Fritsch+Carlson:1980}
\begin{equation}
    f_i = g(x_i)
\end{equation}
for all spline knots indexed by $i \in \{1,K-1\}$,
where in each subinterval $I_i = [x_i, x_{i+1}]$, the function $g(x)$ is a monotonic cubic polynomial defined by
\begin{equation}
    g(x) = f_i H_1(x) + f_{i+1} H_2(x) + d_i H_3(x) + d_{i+1} H_4(x)
\end{equation}
where $d_j = p'(x_j)$ for $j = i$ and $i+1$, and $H_k(x)$ are the usual cubic Hermite basis functions for the interval $I_i$:
\begin{align}
    H_1(x) &= \phi\left(\frac{x_{i+1} - x}{x_{i+1} - x_i}\right)\\
    H_2(x) &= \phi\left(\frac{x - x_i}{x_{i+1} - x_i}\right)\\
    H_3(x) &= -(x_{i+1} - x_i) \, \psi\left(\frac{x_{i+1} - x}{x_{i+1} - x_i}\right)\\
    H_4(x) &= (x_{i+1} - x_i) \, \psi\left(\frac{x - x_i}{x_{i+1} - x_i}\right)\\
\end{align}
where 
\begin{align}
    \phi(t) &= 3 t^2 - 2 t^3 \\
    \psi(t) &= t^3 - t^2
\end{align}

The advantage of utilizing a spline model is the flexibility of modeling a potentially complicated function with a finite number of spline knots. The fact that the spline curves are fully defined by their knot positions and heights streamlines the modeling process considerably.
In particular, the PCHIP model was chosen in this study as it is shape-preserving in the sense that it conserves monotonicity, which controls overshoots and oscillations better in cases where the data are not smooth.  However, it should be noted that cubic splines, while slightly wilder in terms of oscillations, tend to produce more accurate results if the data are smooth (continuous first derivative).

The primary benefit of using spline interpolation is that it allows for the modeling of complicated continuous and smooth functions with a finite number of points~\cite{Ahlberg:1967}.  However, while spline curves are straightforward for one-dimensional data, the required number of spline curves increases exponentially with the dimensionality of the parameter space, making it essential to validate this approach in one dimension before generalizing it further.

\section{Other Potential Approaches}\label{sec4}
Given the challenges of dimensionality, it is tempting to consider simpler approaches.  Interpolation techniques such as linear interpolation and logarithmic interpolation offer alternative approaches to estimating values between known data points. Linear interpolation connects adjacent data points with straight lines, providing a simple and computationally efficient method represented by the formula~\cite{Burden+Faires:2004}
\begin{equation}
    y(x) = y_1 + \frac{y_2-y_1}{x_2 - x_1} (x - x_1)
\end{equation}
where $(x_1, y_1)$ and $(x_2, y_2)$ are the coordinates of the known data points, and $x$ is the point of interest between $x_1$ and $x_2$~\cite{Dorn+etal:2006}.

Conversely, logarithmic interpolation assumes a logarithmic relationship between data points, which is useful when the dependent variable changes exponentially relative to the independent variable~\cite{Burden+Faires:2004}. Logarithmic interpolation is represented by~\cite{Dorn+etal:2006}
\begin{equation}
    y(x) = y_1 \left(\frac{x}{x_1}\right)^{{\log{\frac{y_2}{y_1}}}/{\log{\frac{x_2}{x_1}}}}
\end{equation}
where $(x_1,y_1)$ and $(x_2,y_2)$ are the coordinates of the known data points, and $x$ is the point of interest between $x_1$ and $x_2$~\cite{Ahlberg:1967, Burden+Faires:2004}.
Neither of these models is as flexible as the cubic spline model.

We examined other techniques---for example, multivariate regression models~\cite{Lux+etal:2021}, multivariate interpolation~\cite{Lux+etal:2018, Schaback:2005}, and the linear Shepherd method of linear interpolation~\cite{Thacker+etal:2010}---and found that they are not as convenient for this problem. Below, we discuss their limitations. 

Multivariate regression models the relationship between multiple dependent and independent variables by fitting a linear equation to the observed data~\cite{Lux+etal:2021}. With it, there is a risk of overfitting with many predictors, reducing the model's generalizability to new data, in addition to the fact that a line will fail to capture the non-linear relationships that are common in such data~\cite{Lux+etal:2021}.

Multivariate interpolation~\cite{Lux+etal:2018, Schaback:2005} is a method used to estimate unknown values at specific points based on known values at surrounding points in multiple dimensions. It creates a smooth surface that passes through the given data points, making it useful for filling in gaps in spatial or temporal datasets~\cite{Lux+etal:2018, Schaback:2005}. It also ensures an exact fit to the observed data points and can be applied to both regular and irregular grids of data. However, the method can suffer from overfitting to the noise in the observations, leading to a poor performance on new data, and it can become computationally intensive and less effective with very large datasets~\cite{Lux+etal:2018, Schaback:2005}. Moreover, it may experience boundary issues, where the accuracy of interpolation decreases near the edges of the dataset~\cite{Lux+etal:2018, Schaback:2005}.

The linear Shepherd method of linear interpolation~\cite{Thacker+etal:2010} is a technique that estimates unknown values by taking the weighted average of nearby known values, where the weights are inversely proportional to the distance from the unknown point to the known points, making it a simple and efficient method for local interpolation in multiple dimensions. However, as a linear method, it may not capture non-linear relationships well, which are often present in such data, and it can suffer from boundary effects. Additionally, the interpolation accuracy may not be as high as that of other, more sophisticated techniques, especially in regions with sparse data~\cite{Thacker+etal:2010}.

\section{Methods}\label{sec5}
Ideally, a planet, and its atmospheric absorption spectrum, is defined by at least 30~parameters; further predicting the climate (the ultimate goal) is not just dependent on the existence of these parameters but also the right factorization of each inter-correlation amongst the parameters~\cite{LUVOIR:2019}.~Hence, the ultimate aim of this research is to develop a model which could describe each of the 20 spectral bins in a 30-dimensional space.
As a demonstration of concept in this present work, we focus on a much simpler problem using synthetic data.  Here, we create synthetic spectra with six spectral bins, each of which has a height, $F_b$, that varies with respect to a single planetary parameter, $x$, according to the following functions:
\begin{align}
    F_1(x) &= 0.5 x^2\\\
    F_2(x) &= 0.3 \sin{(1.5 \pi x)} + 0.5\\
    F_3(x) &= 0.2 \cos{(3 \pi x)} + 0.6\\
    F_4(x) &= 0.25 (x + 0.5)^{-2}\\
    F_5(x) &= 0.4 \cos{(\pi x)} + 0.1 x + 0.8\\
    F_6(x) &= 0.1 + 0.4 x
\end{align}

\begin{figure}[H]
 \includegraphics[width=0.7\columnwidth]{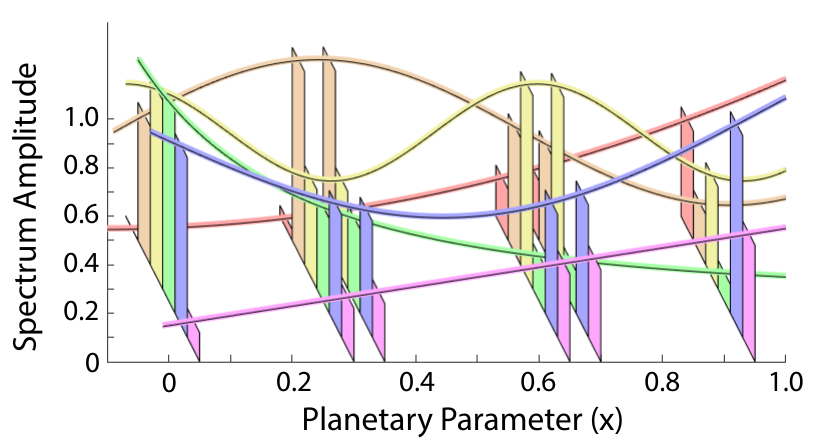}

\caption{An illustration of the synthetic spectrum.  There are six synthetic spectral bins at x-values: $x = \{0.05, 0.30, 0.35, 0.65, 0.70, 0.95\}$.  The spectrum is represented by six spectral bins with amplitudes defined by the functions illustrated by the corresponding curves.\label{fig-3}}
\end{figure}

These six spectral bins are generated as data at a planetary parameter value that we simply call $x$. The x-values of the spectral data are given by $x = \{0.05, 0.30, 0.35, 0.65, 0.70, 0.95\}$, as illustrated in Figure \ref{fig-3}.  No noise was introduced into the synthetic data in this initial study.

Modeling these spectral bins in one dimension is essentially a problem of fitting six curves to the data (one curve per spectral bin). Each curve was modeled as a Piecewise Cubic Hermite Interpolating Polynomial (PCHIP)~\cite{Fritsch+Carlson:1980} defined by six spline knots at positions $\{(x_1,y_1 ),(x_2,y_2 ), \ldots, (x_6,y_6 )\}$, for which both the x and y values were determined using the data.

The PCHIP model for each bin is defined by six to-be-determined PCHIP spline knots.  The $x$-positions of two of the spline knots are held at the extreme points, $x_1 = 0$ and $x_6 = 1$, to limit the polynomial divergences to infinity, whereas the $x$-positions of the other four PCHIP spline knots, $x_2, x_3, x_4$ and $x_5$, are allowed to vary between 0 and 1 with the restriction that they are not allowed to be closer to one another than $\Delta x = 0.1$.  In addition, the resulting PCHIP curves are all constrained to have $y \in [0,1]$ for all $x \in [0,1]$.  The $x$-positions of the four internal PCHIP knots and the $y$-positions of the six PCHIP knots are found separately for each bin by sampling from the posterior probability using the nested sampling algorithm~\cite{Skilling:2006, Skilling:2012}.  The resulting solutions are found by computing the mean of the sampled PCHIP functions.

To perform the posterior sampling using nested sampling, a Gaussian likelihood is employed, which depends on the sum of the square differences between the synthetic spectral bin heights, $y_b(x_i)$, at positions $x_i$, and the value of the PCHIP curve for that bin at $x_i$, $S_b(x_i,{x_{b,k}, y_{b,k}})$, which is defined by the set of spline knots $\{x_{b,k}, y_{b,k}\}$, where $b \in \{1, 2,\ldots, 6\}$ indexes the bin number, and $k \in \{1, 2, \ldots, 6\}$ indexes the specific spline knot (with the values $x_{b,k=1}=0$ and $x_{b,k=6}=1$, fixed for all bins $b$).  The log likelihood is then given by
\begin{equation}
\log{\mathrm{Prob}\big(\{y_b(x_i)\}\big)} = -\sum_{i = 1}^{N}{\frac{\big(y_b(x_i) - S_b(x_i,\{x_{b,k},y_{b,k}\})\big)^2}{2 \sigma^2} - \log{(\sqrt{2 \pi} \sigma)}},
\end{equation}
where the standard deviation is set to $\sigma = 0.001$.

\section{Results}\label{sec6}
\begin{figure}[H]

\includegraphics[width=0.75\columnwidth]{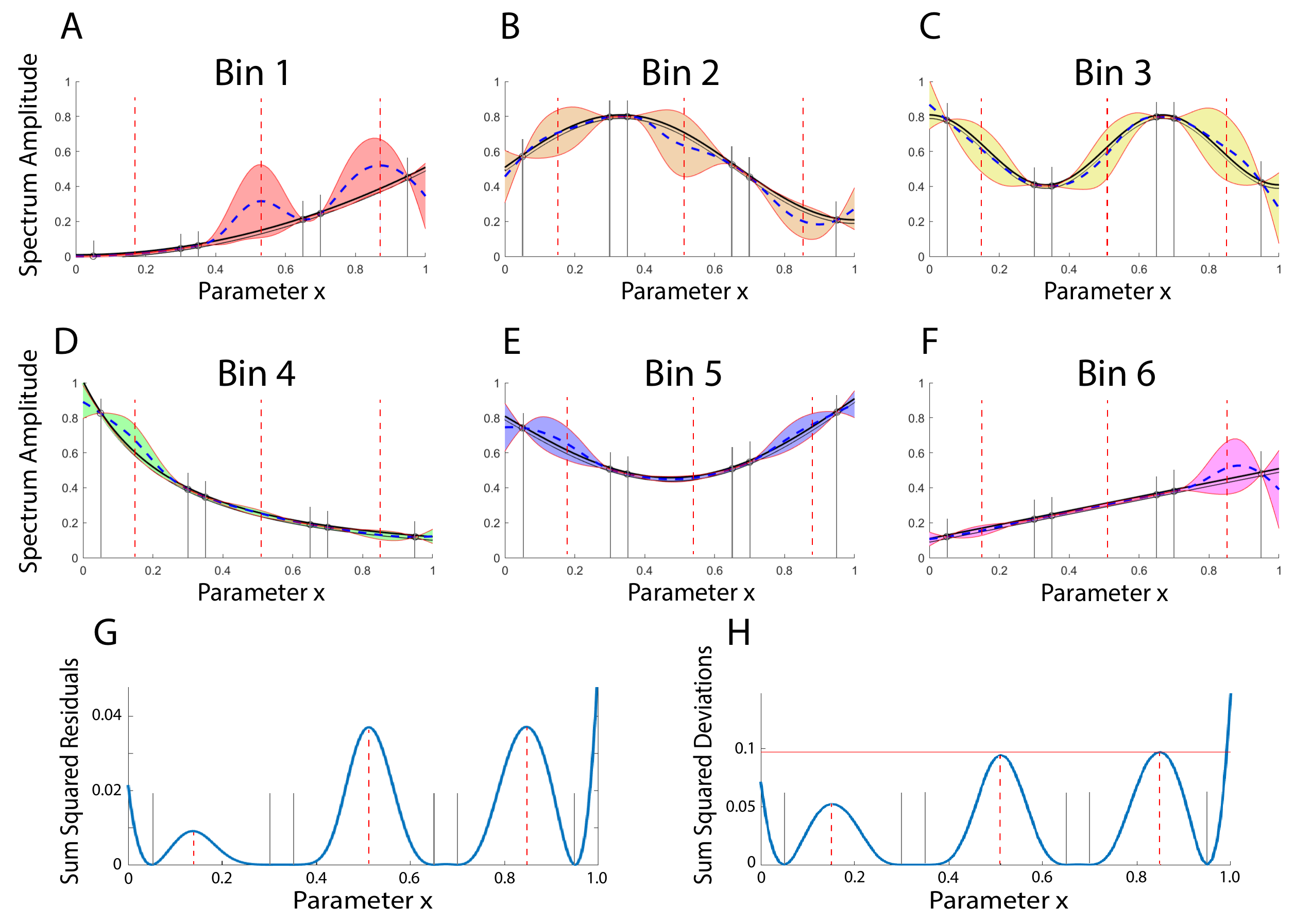}

\caption{(\textbf{A}--\textbf{F}). An illustration of the estimated mean functions (dashed curves) and standard deviations (colored/shaded regions) for the six spectral bins.  The (truth) function from which the data were generated is shown as the solid line curve (see Figure \ref{fig-3}).  The solid vertical lines indicate the positions of the data. The red vertical dashed lines illustrate the locations of greatest uncertainty, which indicate where further spectral measurements would be most informative. (\textbf{G}). An illustration the sum of the squared residuals between the estimated and true functional relationship.  (\textbf{H}). An illustration of the sum of the squared differences (summed over bins) between the estimated functions and the true functions.  Larger values indicate regions of (overall) greater uncertainty proportional to the information that is to be gained by obtaining data there.\label{fig-4}}
\end{figure}

Figure \ref{fig-4}A--F depict the plots of the mean estimated spectral functions (dashed curve) for each bin estimated using the six synthetic spectra at $x = \{0.15, 0.30, 0.45, 0.60, 0.75, 0.90\}$, as illustrated in Figure \ref{fig-3}.  The mean estimated spectral interpolants are indicated by the dashed curve, which is seen to be close to the (true) curves used to generate the data (Figure \ref{fig-3}).  The standard deviations of the estimated interpolants are illustrated by the colored/shaded regions.  These shaded regions indicate the predictive distribution. That is, the shaded regions with the mean denoted by the dashed curve summarize what is known about the functional relationship between the planetary parameter $x$ and the planetary spectra. The most informative regions for sampling data are given by the regions for which the entropy of the predictive distribution is maximum. These are the regions for which sampling will provide the most information. Such informed sampling is known as Bayesian Adaptive Exploration~\cite{Knuth:2003,Loredo:2004,Malakar+Knuth:2011}. The posterior probabilities of the spline knot positions in $x$ and $y$ are what is predicted, and these posterior distributions are Gaussian. The predictive distribution of the spline curve is not Gaussian. In the case of a Gaussian distribution, the entropy is proportional to the variance. So, often, the variance in the predictive distribution, which contributes to the spread of the distribution, is a relatively good approximate measure of the entropy. As a result, the location of the greatest entropy in the interior of the domain of $x$ is given by the peak at $x = 0.85$ (Figure \ref{fig-4}G). It will therefore be most informative to collect another data sample at $x = 0.85$.

Figure \ref{fig-4}H illustrates the sum of the squared residuals, which represents the overall difference between the predicted spectra and the correct solution.
Figure \ref{fig-4}G illustrates the combined uncertainties in the predictive distribution, as a function of the parameter x, found by summing the squared standard deviations (variance) of the mean interpolants in each of the bins.  This plot summarizes where information is most needed.
While the greatest uncertainty remains at the $x = 1$ end point, there are three regions between the data points (at $x = 0.15$, $x = 0.51$, and $x = 0.85$) at which the uncertainties are high and where the most information is to be gained.

\begin{figure}[H]

\includegraphics[width=0.7\columnwidth]{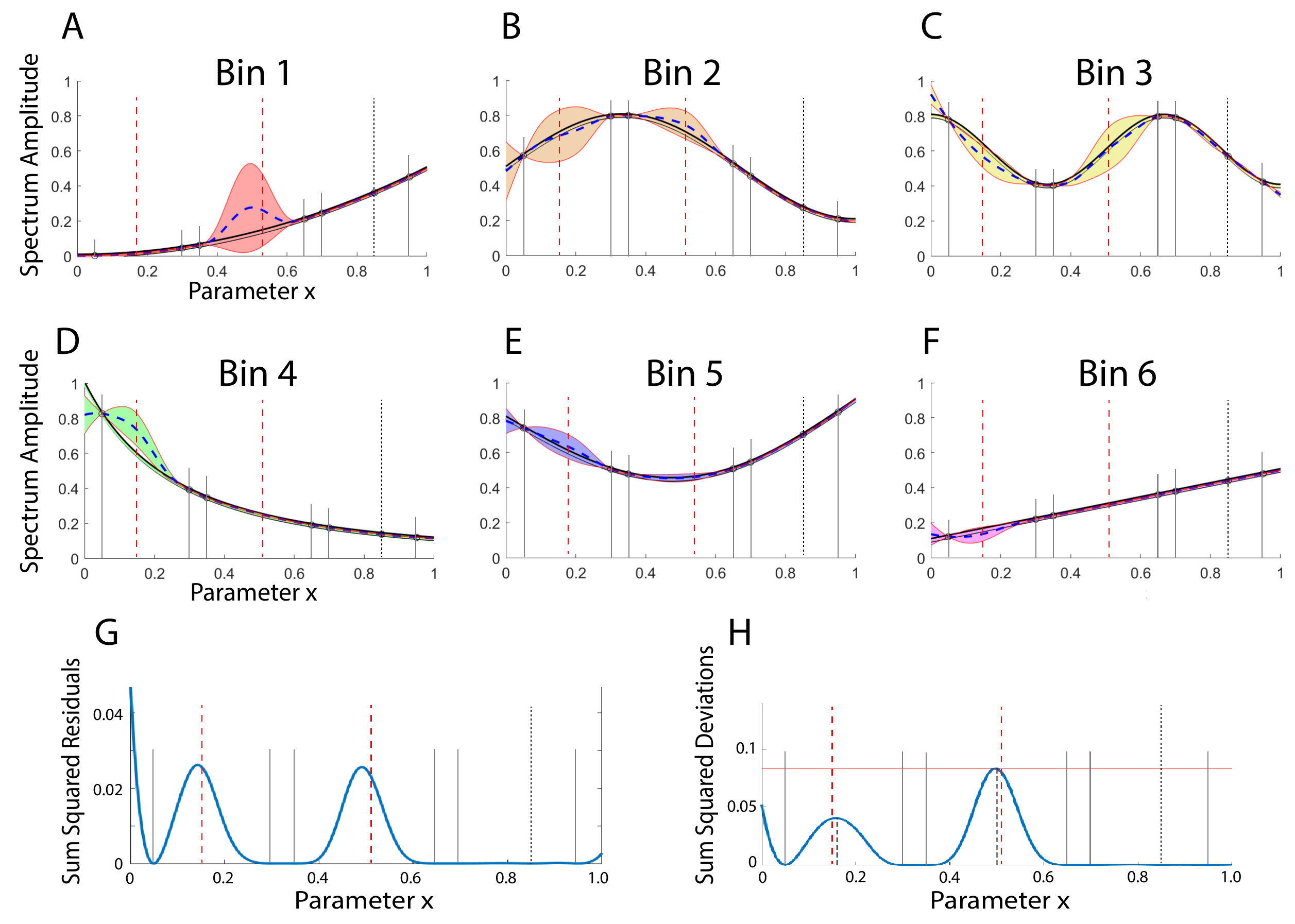}

\caption{(\textbf{A}--\textbf{F}). An illustration of the estimated mean functions (dashed lines) and standard deviations (colored/shaded regions) for the six spectral bins with an additional data spectrum at the location $x = 0.85$, indicated by the black vertical dotted line.  (\textbf{G}) illustrates that the fit has been improved in the region where $x > 0.85$. (\textbf{H}). The sum of the squared deviations shows that collecting data near $x = 0.15$ and $x = 0.51$ would still be most informative, despite the fact the shape and positions of these peaks have shifted slightly, as indicated by the fact that the red vertical dashed lines (indicating the initial positions expected to be most informative) have now slightly deviated from the peaks of the sum squared deviations. \label{fig-5}}  
\end{figure}

Figure \ref{fig-5} illustrates the improvements made to the model by including another data spectrum at $x = 0.85$. Note that not only has the uncertainty on the right-hand side of the domain been brought under control but the uncertainty at the $x = 1.0$ boundary also has. It is important to note that the shapes and positions of the remaining peaks in Figure \ref{fig-5}G,H have shifted slightly.  This highlights the fact that the new information gained by sampling at x = 0.85 has changed the overall landscape. Despite this, sampling data at the original positions of x = 0.15 and 0.51 will still be useful.

\begin{figure}[H]

\includegraphics[width=0.7\columnwidth]{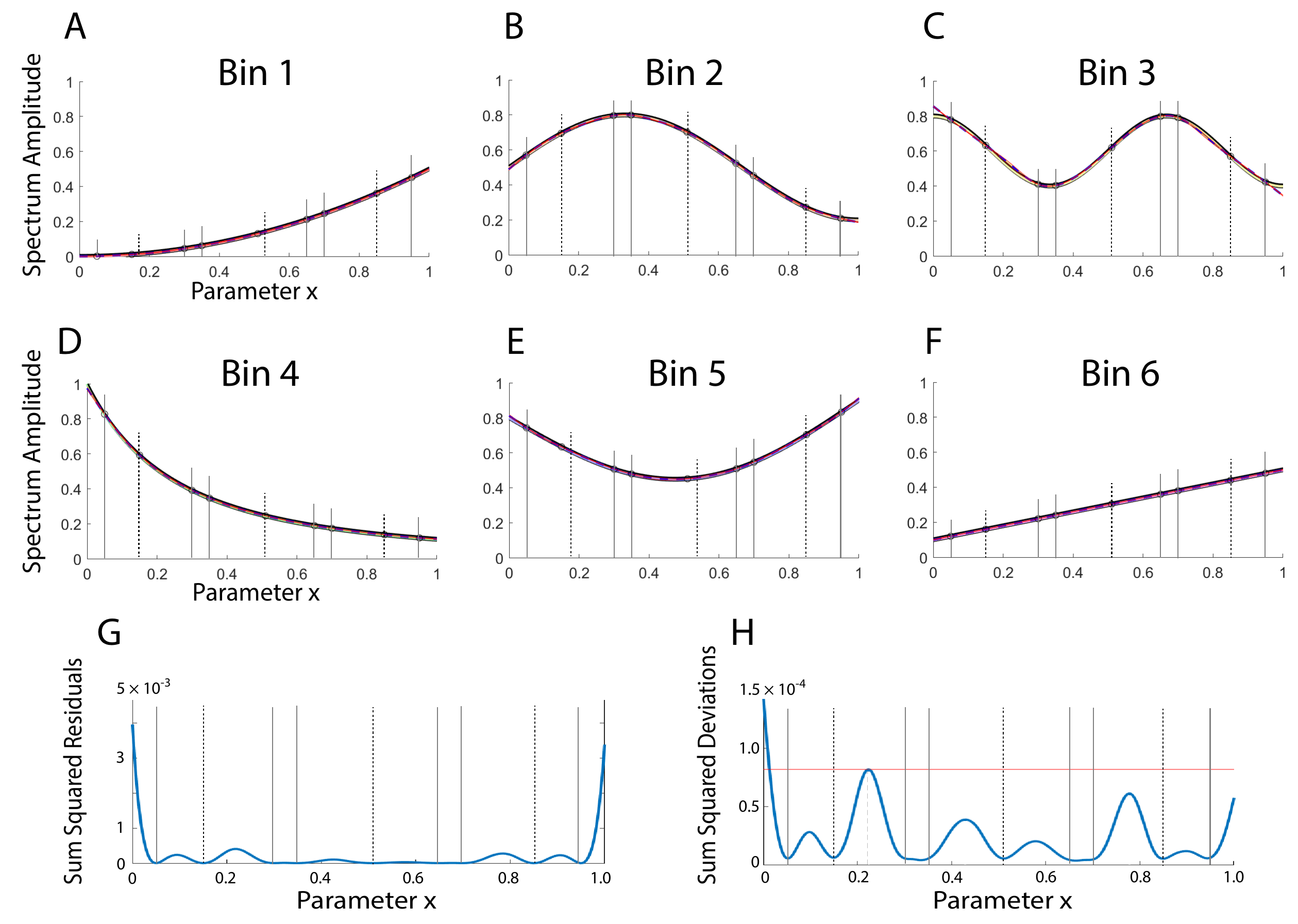}

\caption{(\textbf{A}--\textbf{F}). An
 illustration of the estimated mean functions (dashed curves, which lie under the solid curves) and standard deviations (colored/shaded regions) for the six spectral bins including additional data at all three parameter values of $x = \{0.15, 0.51, 0.85\}$, illustrated by the black vertical dotted lines.  (\textbf{G}). This illustrates that the fit is very good, with the sum of the squared residuals well below $5 \times 10^{-3}$.  (\textbf{H}). The sum of the squared deviations (solid blue curve) shows where collecting additional measurements, if desired, will be most informative.  \label{fig-6}}
\end{figure}

Figure \ref{fig-6} illustrates the utility of including data at all three parameter values of $x = \{0.15, 0.51, 0.85\}$.  Figure \ref{fig-6}G shows that the function is rather well modeled, with the sum of the squared residuals being below $5 \times 10^{-3}$ everywhere. The peaks in the uncertainties in the predictive distribution (Figure \ref{fig-6}H) indicate where collecting more data could improve the model further.

\vspace{-9pt}

\section{Discussion}\label{sec7}
We have demonstrated the basic ideas behind the project to develop a machine-learning-based model of exoplanetary atmospheric absorption spectra based on planetary parameters. This paper presents a proof-of-concept focused on modeling a small number of spectral bins for one planetary parameter in one dimension. Although a PCHIP spline model was convenient to use in this simplified case, it is clear that representing the model as a collection of spline knots in 30 dimensions will be impractical.  This compels us to consider more sophisticated models in future work on the full problem. By considering a simplified problem, we were able to demonstrate the utility of Bayesian Adaptive Exploration, in which sampling the solutions from the posterior and identifying the planetary parameters for which the predictive distribution is the broadest signify the planetary parameter values for which further sampling would be most informative.  This alone is a significant advantage over more simplistic averaging or modeling techniques.

\vspace{6pt} 



\authorcontributions{Conceptualization: K.H.K. and M.J.W.; methodology: V.T., K.H.K., and M.J.W.; software: V.T. and K.H.K.; writing---original draft preparation: V.T.; writing---review and editing: V.T., K.H.K., and M.J.W.; supervision: K.H.K. and M.J.W.; project administration: K.H.K. and M.J.W.; funding acquisition: K.H.K. and M.J.W. All of the authors have read and agreed to the published version of the manuscript.}

\funding{The authors V.T. and K.H.K. were supported by the NASA Interdisciplinary Consortia for Astrobiology Research (ICAR) grant 22-ICAR22\_2-0015. 
M.J.W. was supported by NASA's Nexus for Exoplanet System Science (NExSS) and the NASA Interdisciplinary Consortia for Astrobiology Research (ICAR). M.J.W. also acknowledges support from the GSFC Sellers Exoplanet Environments Collaboration (SEEC), which is funded by the NASA Planetary Science Divisions Internal Scientist Funding Model, and ROCKE-3D, which is funded by the NASA Planetary and Earth Science Divisions Internal Scientist Funding Model.}
\institutionalreview{Not applicable.}
\informedconsent{Not applicable.}
\dataavailability{This study used synthetic data generated using the mathematical functions described in Section \ref{sec4} of this paper.}

\acknowledgments{The authors V.T., K.H.K. and M.J.W. thank the anonymous referee for their detailed review and constructive suggestions, which greatly improved this article.}

\conflictsofinterest{The authors declare no conflicts of interest.}

\begin{adjustwidth}{-\extralength}{0cm}

\reftitle{References}

\PublishersNote{}
\end{adjustwidth}
\end{document}